\documentclass[aps,prd,preprint,superscriptaddress,amsmath,amssymb,showpacs]{revtex4-1}
\usepackage{dcolumn}
\usepackage{graphicx}
\usepackage{float}
\usepackage{physics}
\usepackage{footnote}
\usepackage{ulem}
\usepackage[colorlinks=true,allcolors=blue]{hyperref}

\begin{document}
\title{$\mathcal{R}^2$ curvature-squared corrections on Langevin diffusion coefficients}
\date{\today  \hspace{1ex}}
\author{Qi Zhou}
\email{qizhou@mails.ccnu.edu.cn}
\affiliation{Key Laboratory of Quark \& Lepton Physics (MOE) and Institute of Particle Physics,Central China Normal University, Wuhan 430079, China}
\affiliation{Helmholtz Research Academy Hessen for FAIR (HFHF),GSI
Helmholtz Center for Heavy Ion Physics. Campus Frankfurt, 60438
Frankfurt, Germany}
\affiliation{Institut f\"ur Theoretische Physik, Johann Wolfgang Goethe-Universit\"at,Max-von-Laue-Str.\ 1, D-60438 Frankfurt am Main, Germany}

\author{Ben-Wei Zhang}
\email{bwzhang@mail.ccnu.edu.cn}
\affiliation{Key Laboratory of Quark \& Lepton Physics (MOE) and Institute of Particle Physics,Central China Normal University, Wuhan 430079, China}

\begin{abstract}
The effect of finite coupling corrections to the Langevin diffusion coefficients on a moving heavy quark in the Super Yang-Mills plasma is investigated. 
These corrections are related to curvature squared corrections in the corresponding gravity sector. 
We compare the results of both longitudinal and perpendicular Langevin diffusion coefficients with those in $\mathcal{N}$=4 Super Yang-Mills plasma. It is observed that the curvature-squared corrections influence the Langevin diffusion coefficients, and the corrections for both Langevin diffusion coefficients demonstrate the dependence on the velocity of the moving heavy quark and the specifics of the higher derivative correction.
In addition, we conduct calculations for the Langevin diffusion coefficients of a moving heavy quark within the Gauss-Bonnet background.
\end{abstract}

\maketitle
\section{Introduction}\label{sec1}

The heavy ion collision experiments (HICs) at the Relativistic Heavy Ion Collider (RHIC) and the Large Hadron Collider (LHC) are believed to create almost the most perfect fluid Quark Gluon Plasma (QGP) \cite{BRAHMS:2004adc,PHENIX:2004vcz,PHOBOS:2004zne,STAR:2005gfr}. This provides a novel window for studying the physics of Quantum Chromodynamics~(QCD) at a strongly coupled regime.
Since the properties of a strongly coupled system cannot be reliably calculated directly by perturbative techniques, one has to resort to some nonperturbative approaches to conquer the challenges.

AdS/CFT correspondence \cite{Maldacena:1997re,Witten:1998qj,Gubser:1998bc} is one very promising approach to deal with these problems in QCD in a strongly coupled scenario which cannot be handled properly by perturbative methods \cite{Witten:1998zw,Aharony:1999ti}.
It is believed, on gravity sector, that an external quark on the gauge theory side is related to a string which has a single endpoint at the boundary and extends down to the horizon of an AdS black hole \cite{Gubser:2006bz,Herzog:2006gh}.
Moreover, the diffusion of heavy quarks in a strongly coupled plasma can be understood as the fluctuation correlations of the trailing string.
The study of the stochastic nature of a heavy quark in a holography was proposed by~\cite{Gubser:2006nz,Casalderrey-Solana:2007ahi}. Soon the stochastic motion is formulated as a Langevin process~\cite{deBoer:2008gu,Son:2009vu}. 
Since the heavy quarks in HICs experiments are relativistic in many cases, the relativistic Langevin equation is studied in~\cite{Giecold:2009cg} as well as in non-conformal frameworks in~\cite{Gursoy:2010aa,Kiritsis:2011bw} towards the multiple scales of QCD.

Methods in AdS/CFT relate a 4-dimensional $\mathcal{N} = 4$ Super Yang-Mills (SYM) theory to a type-IIB string theory on the $AdS_5\times S^5$, and string theory contains higher derivative corrections to classical gravity from stringy $1/\lambda$ or quantum effects $1/N_c$ corrections. 
It's natural to conduct computations for finite 't Hooft coupling of gauge theory corresponding to studying the effect of higher derivative corrections on computations in classical Einstein gravity.
The leading order correction in $1/\lambda$ arises from stringy corrections to the low energy effective action of type-IIB supergravity $\alpha^{\prime 3}\mathcal{R}^4$ \cite{Pawelczyk:1998pb}, and apply such corrections to the ratio of shear viscosity over to entropy density of gauge field, $\eta/s$, was calculated in~\cite{Benincasa:2005qc,Buchel:2004di}.
Soon, it was found in \cite{Kats:2007mq} that the universal low bound on the $\eta/s$ and causality can be violated when given a general $\mathcal{R}^2$ corrections to the gravitational action in GB gravity \cite{Brigante:2007nu,Kats:2007mq,Brigante:2008gz,Buchel:2008wy,Neupane:2008dc,Neupane:2009sx,Ge:2008ni}.
For instance the authors of \cite{Buchel:2008wy} studied the ratio of $\eta/s$ in 5 dimension setting, and the authors of \cite{Ge:2008ni} studied the ratio in RN-AdS black branes solution finding that $\eta/s$ bound is violated and the Maxwell charge slightly reduces the deviation. 
For instance, the authors of \cite{Buchel:2008wy} studied the ratio of $\eta/s$ in a 5-dimensional setting, while \cite{Ge:2008ni} discovered that the $\eta/s$ bound is violated and the presence of Maxwell charge slightly reduces the deviation in the RN-AdS black brane solution.
Motivated by these vast string landscapes, the effect of high derivative curvature $\mathcal{R}^4$ or $\mathcal{R}^2$ corrections on different aspects of the properties of QGP has been studied in \cite{Ali-Akbari:2009svi,Ali-Akbari:2010xwz,Zhang:2023kzf,Arnold:2012uc,Atashi:2019mlo,Li:2018lsl,Zhang:2018yzh}. 
Besides, the authors of \cite{Myers:2010ru,Oliva:2010eb,Oliva:2010zd,Zhu:2024ddp} studied curvature-cubic $\mathcal{R}^3$ corrections to $\mathrm{AdS_5}$.

One of the significant applications of the AdS/CFT correspondence is the investigation of jet quenching phenomena involving high transverse momentum partons produced in HICs.
It is found in \cite{Ficnar:2013qxa} that introducing $\mathcal{R}^2$ corrections to the $\mathrm{AdS_5}$ yields a substantial increase in nuclear modification factor $R_{AA}$.
The initial $1/\lambda$ correction to the jet quenching parameter was found in~\cite{Armesto:2006zv}, followed by the jet quenching parameter include $\mathcal{R}^2$ correction \cite{Zhang:2015hkz,BitaghsirFadafan:2010rmb}.
The trailing string, which models the drag force on a moving  heavy quark, has been examined within the framework of higher derivative gravity to investigate the $\mathcal{R}^2$ and $\mathcal{R}^4$ correction to drag force in \cite{Vazquez-Poritz:2008lxn,Fadafan:2008gb}.
In this study, we focus on investigating $\mathcal{R}^2$ corrections to Langevin diffusion coefficients (LGV-coefficients) that are related to the fluctuations of the trailing string.

The organization of this paper is as follows. 
In the subsequent section, section \ref{sec2}, we will review the main procedures to deduce LGV-coefficients within the membrane paradigm. Additionally, we also discuss numerical results of $\mathcal{R}^2$ correction to LGV-Coefficients in section \ref{sec2}.
In section \ref{sec3}, we will study $\mathcal{R}^2$ correction with GB gravity to LGV-coefficients as in section \ref{sec2}. 
The last part section \ref{sec4} is devoted to conclusions and discussion.

\section{Curvature squared corrections to AdS-Schwarzschild black brane on the Langevin diffusion coefficient}\label{sec2}
The curvature squared corrections to solution of $AdS_5$-Schwarzschild black brane can be described by the general action \cite{Brigante:2008gz,Brigante:2007nu}
\begin{equation}
\begin{split}
\label{action}
S = \frac{1}{16\pi G_5}&\int d^5 \sqrt{-g}\times
  \left[\mathcal{R}-\Lambda+L^2\left(c_1 \mathcal{R}^2+c_2\mathcal{R}_{\mu\nu}\mathcal{R}^{\mu\nu}+c_3 \mathcal{R}_{\mu\nu\rho\sigma}\mathcal{R}^{\mu\nu\rho\sigma}\right)\right]
\end{split}  
\end{equation}
where $G_5=\pi L^3/2N_c^2$ is 5-dimensional Newton constant, $\mathcal{R}$ is the Ricci scalar, $\mathcal{R}_{\mu\nu}$ and $\mathcal{R}_{\mu\nu\rho\sigma}$ are the Ricci tensor and Riemann tensor, respectively. 
The negative cosmological constant $\Lambda=-\frac{12}{L^2}$ creates an $AdS$ space with the radius $L$. 
The parameter $c_i$ are expected to be of $o(\alpha')$ which means $c_i=0$ in the limit of large 't Hooft coupling ($\lambda\rightarrow\infty$). 
The shear viscosity to entropy ratio, was found in \cite{Kats:2007mq,Noronha:2009ia} to be $\frac{\eta}{s}=\frac{1}{4\pi}(1-8c_3)+\mathcal{O}(c_i^2)$, and the viscosity bound is violated when $c_3 > 0$.

The black brane solution of $AdS_5$ space for (\ref{action}) is given by \cite{Kats:2007mq},
\begin{eqnarray}
\label{metric-r2}
ds^2=-(\frac{r^2}{L^2})f(r)dt^2+(\frac{r^2}{L^2})d\Vec{x}^2+\frac{L^2}{r^2f(r)}dr^2,
\end{eqnarray}
where
\begin{equation}
 f(r)=1-\frac{r_0^4}{r^4}+a+b \frac{r_0^8}{r^8}
\end{equation}
and
\begin{equation}
a=\frac{2}{3} \left( 10c_1 + 2c_2
+c_3 \right),\,\,\, b=2c_3.
\label{alpha}
\end{equation}
The boundary of the asymptotically AdS geometry is located at $r\rightarrow\infty$ where $r$ denotes the 5th dimensional radial coordinate, and $(t, \vec{x})$ label the left 4-dimensional spacetime of gauge theory on the boundary. 
One can solve $f(r_h)=0$ to find location of the horizon $r = r_h$, where $r_h$ depends on $a$, $b$ and $r_0$. 
The heat bath temperature is given by
\begin{equation}
T_{R^2}=\frac{r_0}{\pi L^2} \left( 1+ \frac{1}{4} a - \frac{5}{4}
b\right)\label{TR2},
\end{equation} 
where $r_0$ depends on both $a$ and $b$ for a fixed temperature $T_{R^2}$.

By following the authors of~\cite{Gursoy:2010aa,Giataganas:2013hwa,Finazzo:2016mhm}, we compute the LGV-coefficients of a heavy quark in squared-curvature correction background. 
It more conveniences to conduct the calculation in a more general form,
\begin{equation}
    ds^2=g_{tt}dt^2+g_{ii}dx_i^2+g_{rr}dr^2.
\end{equation}
From (\ref{metric-r2}), one has the 
\begin{equation}
\label{metric-r3}
    g_{tt}=-(\frac{r^2}{L^2})f(r), \quad g_{ii}=\frac{r^2}{L^2}, \quad g_{rr}=\frac{L^2}{r^2f(r)}.
\end{equation}

Holographically, the moving heavy quark of infinite mass on the boundary CFT correspond to the endpoint of the trailing string. The string dynamics are captured by the Nambu-Goto action
\begin{equation}
\begin{split}
    \label{nbaction}
     S_{NG}=-\frac{1}{2\pi\alpha'}\int \mathrm{d}\tau \mathrm{d}\sigma \sqrt{-\mathrm{det} \gamma_{\alpha\beta}}, \qquad      \gamma_{\alpha\beta}=g_{\mu\nu}\partial_\alpha X^{\mu} \partial_\beta X^{\nu}.
\end{split}
\end{equation}
where $\gamma_{\alpha\beta}$ is the induced metric, and $g_{\mu\nu}$ and $X^{\mu}$ are the branes metric and target space coordinates.

Given a moving heavy quark with a constant velocity $v$ on the boundary along the chosen direction $x_{p}(x_p=x,y,z)$, one can choose to compute in static gauge for the string world-sheet and has the usual parametrization
\begin{equation}
    \label{gauge}
    t=\tau, \quad r=\sigma,\quad x=vt+\xi(r),
\end{equation}
where $\xi$ is the profile of the string in the bulk.
It is deduced the world-sheet metric
\begin{equation}
\label{metric-ws}
    \gamma_{\alpha\beta}=
\left(
        \begin{matrix}
            g_{tt}+v^2g_{pp} &  g_{pp}v\xi' \\
             g_{pp}v\xi' & g_{rr}+g_{pp}\xi^{'2}
        \end{matrix}
\right),
\end{equation}
and the corresponding action
\begin{equation}
    \label{nbaction}
     S_{NG}=-\frac{1}{2\pi\alpha'}\int \mathrm{d}t \mathrm{d}r \sqrt{-(g_{tt}g_{rr}+g_{tt}g_{pp}\xi'^2+g_{pp}g_{rr}v^2)}.
\end{equation}
The $g_{pp}$ is the corresponding metric component in the $x_{p}$ direction.
It's obvious that radial conjugate momentum $\pi_{\xi}$ is conserved for the simple motion
\begin{equation}
\label{canonicalmomentum}
    \begin{split}
        \pi_{\xi}=\frac{\delta S}{\delta \xi}=-\frac{1}{2\pi\alpha'}\frac{g_{tt}g_{pp}  \xi'}{2\sqrt{-(g_{tt}g_{rr}+g_{tt}g_{pp}\xi'^2+g_{pp}g_{rr}v^2)}}.
    \end{split}
\end{equation}
It is easy to find $\xi'$ from (\ref{canonicalmomentum}) as
\begin{equation}
    \xi'=\sqrt{\frac{-g_{tt}g_{rr}-g_{pp}g_{rr}v^2}{g_{tt}g_{xx}(1+\frac{g_{tt}g_{xx}}{C^2})}},
\end{equation}
where $C\equiv 2\pi\alpha'\pi_{\xi}$. The world-sheet of the string has a horizon and turns out to be the same with critical point $r_c$ at which both numerator and denominator change their sign.
By inserting (\ref{metric-r3}) into $\gamma_{\alpha\alpha}(r_c) = 0$, one can identify the critical point $r_c$ as
\begin{equation}
\label{criticalr}
    r_c=\sqrt[4]{\frac{r_0^4 \sqrt{1-4 b \left(a-v^2+1\right)}+r_0^4}{2 \left(a-v^2+1\right)}}.
\end{equation}

One can also find the effective temperature $T_{ws}$ of the world-sheet horizon by diagonalizing world-sheet metric (\ref{metric-ws}). One can change coordinates to diagonalize the induced metric, by means of the reparametrization,
\begin{equation}
    d\tau\rightarrow d\tau-\frac{\gamma_{\alpha\beta}}{\gamma_{\alpha\alpha}}d\sigma.
\end{equation}
The diagonal induced world-sheet metric $h_{\alpha\beta}$ given
\begin{equation}
    h_{\alpha\beta}=
\left(
        \begin{matrix}
            g_{tt}+g_{pp}v^2 &    \\
               & \frac{g_{tt}g_{pp}g_{rr}}{g_{tt}g_{pp}+ \left(2\pi\alpha'\pi_{\xi} \right)^2}
        \end{matrix}
\right).
\end{equation}
Following the usual procedure, the effective world sheet temperature reads
\begin{equation}
\begin{aligned}
    T^{2}_{ws}&=\frac{1}{16\pi^2}\left(h_{\alpha\alpha}'\,\left(h^{\beta\beta}\right)'\right)\bigg|_{r_c}\\
    &=\frac{1}{16\pi^2}\left[(g_{tt}+v^2g_{pp})'(\frac{g_{tt}g_{pp}+v^2(g_{pp}|_{r_c})^2}{g_{tt}g_{pp}g_{uu}})'\right]^2\bigg|_{r=r_{c}}\\
    &=\frac{1}{16\pi^2}\abs{\frac{g^{'2}_{tt}-v^4g^{'2}_{pp}}{g_{tt}g_{pp}}}\bigg|_{r=r_{c}}\\
    &=\frac{1}{16\pi^2}\abs{\frac{1}{g_{tt}g_{rr}}(g_{tt}g_{pp})'(\frac{g_{tt}}{g_{pp}})'}\bigg|_{r=r_{c}}.
    \label{tws}
\end{aligned}
\end{equation}
By inserting (\ref{criticalr}) into (\ref{tws}), one has the effective world sheet temperature $T^{ws}_{R^2}$ as 
\begin{equation}
\begin{aligned}
\label{tws1}
    T^{ws}_{R^2}=\frac{1}{4\pi}\sqrt{\abs{\left(\frac{8 b r_0^8}{r_c^9}-\frac{4 r_0^4}{r_c^5}\right) \left(4 r_c^3 \left(a+\frac{b r_0^8}{r_c^8}-\frac{r_0^4}{r_c^4}+1\right)+r_c^4 \left(\frac{4 r_0^4}{r_c^5}-\frac{8 b r_0^8}{r_c^9}\right)\right)}}.
\end{aligned}
\end{equation}
In the conformal limit, where $a\rightarrow 0,\,b\rightarrow 0$, the background solution reduces to AdS-BH, and the world-sheet temperature (\ref{tws1}) is simply related to bulk temperature 
\begin{equation}
    \lim_{a\rightarrow 0,b\rightarrow 0}T^{ws}_{R2}=\frac{T_{\text{SYM}}}{\sqrt{\gamma_v}},
\end{equation}
where $\gamma_v$ is the Lorentz factor $\gamma_v=1/\sqrt{1-v^2}$ and $T_{\text{SYM}}$ is the bulk temperature in conformal limit.

Considering the fluctuation in classical trailing string, one has 
\begin{equation}
    t=\tau,\quad r=\sigma,\quad x_{p}=vt+\xi(\sigma)+\delta x_p(\tau,\sigma),
\end{equation}
where the fluctuation taking the form $\delta x_p(\tau,\sigma)$ along and transverse to the direction of $x_p$. A simple expression for the quadratic action in the worldsheet embedding fluctuations capturing fluctuations of heavy quark reads
\begin{equation}
\begin{split}
\label{sfluctuation}
    S_2=-\frac{1}{2\pi\alpha'}\int d\tau d\sigma \frac{H^{\alpha\beta}}{2}
    \times(N[r]\partial_{\alpha}\delta x_p\partial_{\beta}\delta x_p+\sum_{i\neq p} g_{ii}\partial_{\alpha}\delta x_i\partial_{\beta}\delta x_i),\\
    N(r)\equiv\frac{g_{tt}g_{pp}+C^2}{g_{tt}+g_{pp}v^2},\\
    H^{\alpha\beta}=\sqrt{-det(h)}h^{\alpha\beta},\\
    \end{split}
\end{equation}
where $h^{\alpha\beta}$ is inverse of the diagonalized induced world-sheet metric. 
We recommend referring to \cite{Gubser:2006nz,Gursoy:2010aa,Giataganas:2013hwa,Chakrabortty:2013kra} for a detailed proof.
For an arbitrary massless fluctuation $\phi$ with an action
\begin{equation}
\label{examplefluctuation}
    S_2=-1/2\int dxdr\sqrt{-g}Q(r)g^{\alpha\beta}\partial_{\alpha}\partial_{\beta}\phi.
\end{equation}
Taking advantage of the membrane paradigm~\cite{Iqbal:2008by}, one can directly reads the transport coefficient associated with the retarded Green’s function form (\ref{examplefluctuation}) without solving motion equation as
\begin{equation}
\begin{split}
\label{membraneparadigm}
    \chi_{R}&=-\lim_{k_{\mu}\rightarrow0}\frac{\Im G_{R}(\omega,\vec{k})}{\omega}=Q(r_h)
    \end{split}
\end{equation}
where $Q$ is the only effective coupling of the fluctuation and the metric dependence drops out in 2-dimension worldsheet black hole horizon. 

For sufficiently large times, the temporal correlation functions of the random force operator on a Brownian particle are proportional to Dirac delta distributions, with the proportionality factors defining the Langevin diffusion coefficients. On the other hand, the noise term is determined by the symmetrized real-time correlation functions of the random forces over the statistical ensemble. Then the LGV-coefficient can be defined in terms of the symmetric correlator $G_{sym}$ \cite{Gubser:2006nz} as
\begin{equation}
\begin{split}
\label{lgv_def}
    \kappa_{d}&=\lim_{\omega\rightarrow0} G^{d}_{sym}(\omega)\\
    &=-\coth{\frac{\omega}{2T_{ws}}}\lim_{\omega\rightarrow0} (\Im G^d_{R}(\omega))\\
    &=-2T_{ws}\lim_{\omega\rightarrow0} \frac{\Im G^d_{R}(\omega)}{\omega}\\
    &=2T_{ws}\chi_{R}^{d}\\
&=2T_{ws}Q^{d}(r_c).
\end{split}
\end{equation}
where $d=(\perp,\parallel)$. The second step has employed the $\omega\rightarrow 0$ limit of $G^{d}_{sym}(\omega)=\coth{\frac{\omega}{2T}}\Im G^{d}_{R}(\omega)$ \cite{Casalderrey-Solana:2007ahi}, and the third step has employed (\ref{membraneparadigm}).
By comparison of (\ref{sfluctuation}) and (\ref{examplefluctuation}), one yields,
\begin{equation}
\label{mp_def}
        Q^{\perp}=\frac{1}{2\pi\alpha}g_{kk}\bigg|_{r=r_c},\qquad
        Q^{\parallel}=\frac{1}{2\pi\alpha}\lim_{r\rightarrow r_c}N(r)=\frac{1}{2\pi\alpha}\frac{(g_{tt}g_{pp})'}{g_{pp}(\frac{g_{tt}}{g_{pp}})'}\bigg|_{r=r_c}.
\end{equation}
Noticing $N(r_c)=\frac{0}{0}$, therefore one needs to use the L’Hopital’s rule to calculate the limit. One can also insert (\ref{mp_def}) into (\ref{lgv_def}) and 
it yield,
\begin{equation}
\label{kappa}
\kappa_{\perp}= \frac{1}{\pi \alpha'}g_{kk}|_{r=r_c}T_{ws},\qquad \kappa_{\parallel}=\frac{1}{\pi\alpha'}
\frac{( g_{tt}g_{pp})'}{g_{pp} (\frac{g_{tt}}{g_{pp}})'}\bigg|_{r=r_c}T_{ws}.
\end{equation}

In our case, one can also insert metric (\ref{metric-r3}) into (\ref{kappa}) and yield,
\begin{equation}
\label{kappat}
\kappa_{\perp}=\frac{\sqrt{\lambda}}{\pi}r_c^2T^{ws}_{R2},
\end{equation}
and
\begin{equation}
\label{kappal}
\kappa_{\parallel} =\frac{r_c^2 \sqrt{\lambda} \left(b r_0^8-(a+1) r_c^8\right)}{\pi  r_0^4 \left(2 b r_0^4-r_c^4\right)} T^{ws}_{R2},
\end{equation}
where we used the fact that $\alpha' = \frac{L^2}{\sqrt{\lambda}} = \frac{1}{\sqrt{\lambda}}$.

In conformal limit, one can obtain the well-known results~\cite{Gubser:2006nz,Casalderrey-Solana:2007ahi} by taking limits of both $a \to 0$ and $b \to 0$,
\begin{equation}
\begin{split}
    \kappa_{\perp}^{SYM}=\sqrt{\lambda} \pi T_{\text{SYM}}^3 \gamma_v^{\frac{1}{2}},\qquad
    \kappa_{\parallel}^{SYM}=\sqrt{\lambda} \pi T_{\text{SYM}}^3 \gamma_v^{\frac{5}{2}}.
    \label{lgvsym}
\end{split}
\end{equation}
One can check that (\ref{kappat}) and (\ref{kappal}) reduce to these results by taking the limit of $\alpha \rightarrow 0$ and $\beta \rightarrow 0$.

\begin{figure}[H]
    \includegraphics[width=8.6cm]{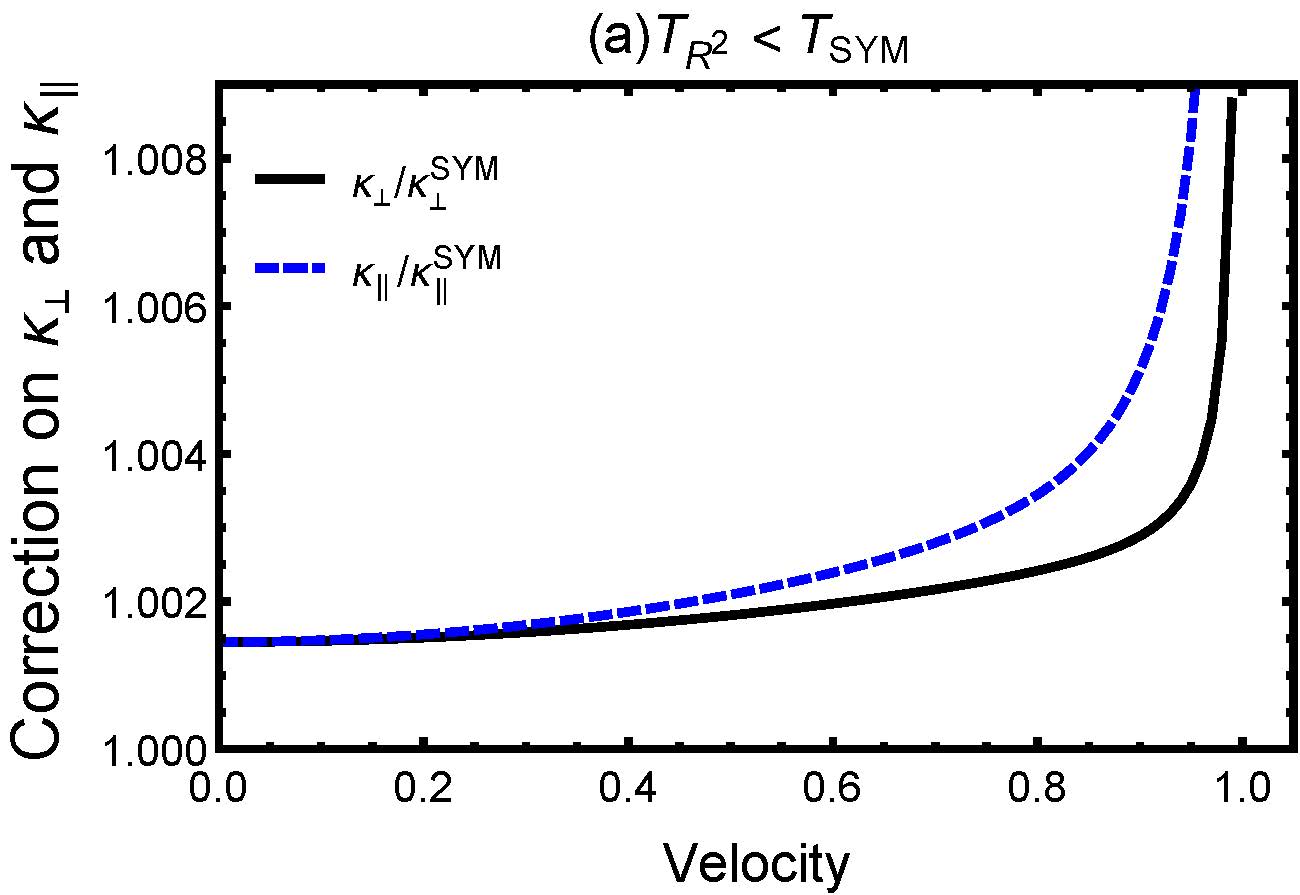}
    \includegraphics[width=8.6cm]{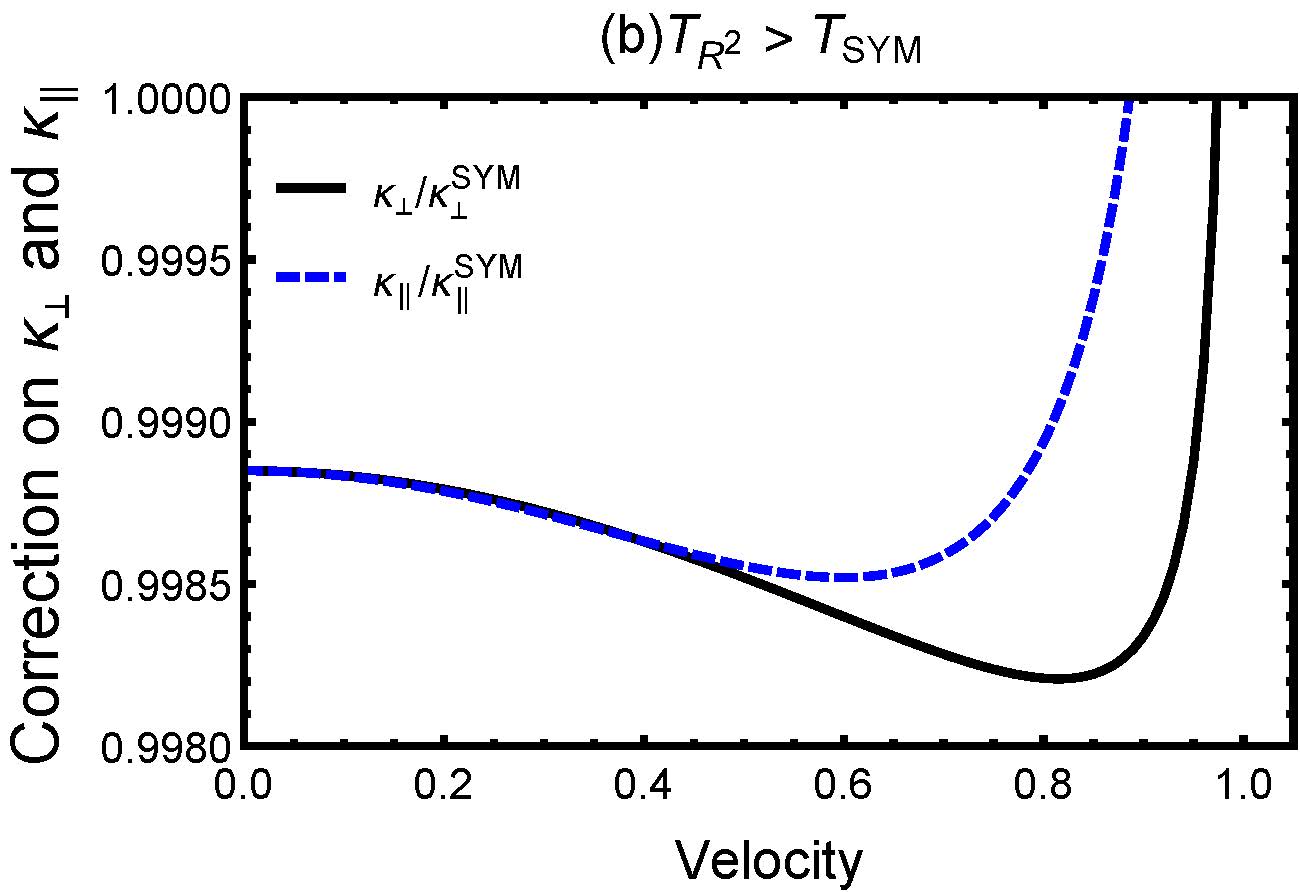} 
    \caption{The correction to transverse LGV-coefficients $\kappa_{\perp}$ and longitudinal LGV-coefficients $\kappa_{\parallel}$ as a function of velocity of the heavy quark, normalized by the conformal limit respectively.}
    \label{figure:R2}
\end{figure}

The effects from $\mathcal{R}^2$ corrections to the classical trailing string, which models the drag force on a moving heavy quark in SYM plasma, were studied by the authors of \cite{Fadafan:2008gb} at two distinct scenarios: $T_{R^2}< T_{SYM}$ and $T_{R^2}> T_{SYM}$. Following their convention, we find it convenient to explore the curvature squared corrections on the fluctuations of the trailing string that is related to LGV-coefficients. More precisely, we investigate the $\mathcal{R}^2$ corrections to $\kappa_{\perp}$ and $\kappa_{\parallel}$ by evaluating (\ref{kappat}) and (\ref{kappal}) using two distinct sets of values for the parameters $a$ and $b$.

Fig.\ref{figure:R2} demonstrates the impact of $\mathcal{R}^2$ corrections on LGV-coefficients where $\kappa_{\perp}$ and $\kappa_{\parallel}$ normalized by the SYM result given in (\ref{lgvsym}), with two scenarios at the same heat bath temperature ($T_{R^2}=T_{SYM}$).
Plots~(a) in Fig.\ref{figure:R2} show $\mathcal{R}^2$ corrections on LGV-coefficients at fixed small values of  $a=-0.0005$ and $b=+0.0006$ corresponding to $T_{R^2}<T_{SYM}$. 
It is clear from this plot that the $\mathcal{R}^2$ corrections to both $\kappa_{\perp}$ and $\kappa_{\parallel}$ are increased monotonically with increasing the moving velocity of the heavy quark and corrections to the LGV-coefficient are larger than SYM case at all velocities.
However, it is also observed that this kind of $\mathcal{R}^2$ correction can also be smaller than $\mathcal{N}=4$ SYM results in Plots~(b) in Fig.\ref{figure:R2} at $a=-0.0005$ and $b=-0.0007$ corresponding to $T_{R^2}>T_{SYM}$.
In this case, a critical velocity ($v_c$) exists where the corrections increase both $\kappa_{\perp}$ and $\kappa_{\parallel}$ if $v>v_c$ while corrections decrease both $\kappa_{\perp}$ and $\kappa_{\parallel}$ if $v>v_c$. 

As a result, we conclude that the finite coupling corrections affect the both $\kappa_{\perp}$ and $\kappa_{\parallel}$ on a moving quark in the strongly-coupled plasma and depend on the details of curvature squared corrections. The LGV-coefficient can be larger than or smaller than that in the infinite-coupling case. 
But at a fixed velocity, one can always find that the corrected $\kappa_{\parallel}$ is at least as large as the corrected $\kappa_{\perp}$.
Moreover, the universal relation, $\kappa_{L}\geq\kappa_{T}$ which founded by the authors of \cite{Gursoy:2010aa}, always hold both when $T_{R^2}>T_{SYM}$ and $T_{R^2}<T_{SYM}$.
Our findings are similar to the case of drag force on a moving heavy quark in \cite{Fadafan:2008gb}.

\section{Gauss-Bonnet gravity background}\label{sec3}

The Gauss-Bonnet~(GB) gravity \cite{Zwiebach:1985uq} is one of the most interesting theories of gravity with curvature squared correction in five dimensions. The exact solutions and thermodynamic properties of the GB background were discussed in~\cite{Cai:2001dz,Nojiri:2001aj,Nojiri:2002qn}.
One can also consider the GB gravity as a special case of the general action (\ref{action}) where $c_2=-4c_1$ and $c_1=c_3=\lambda_{GB}/2$. This yields the action defined as,
\begin{equation}
    \begin{split}
S=&\frac{1}{16\pi G_5}\int d^5 x\sqrt{-g}\times\Bigg[\mathcal{R}-\Lambda+L^2\frac{\lambda_{GB}}{2}\left(\mathcal{R}^2-4\mathcal{R}_{\mu\nu}\mathcal{R}^{\mu\nu}+\mathcal{R}_{\mu\nu\rho\sigma}\mathcal{R}^{\mu\nu\rho\sigma}\right)\Bigg].
    \end{split}
\end{equation}
The dimensionless Gauss-Bonnet coupling constant $\lambda_{GB}$ can be constrained by causality \cite{Brigante:2007nu} and the positive boundary energy density on the boundary \cite{Hofman:2008ar} to be:
\begin{equation}
-\frac{7}{36}<\lambda_{GB}\le \frac{9}{100}.
\end{equation}

A black hole solution in this case is known analytically \cite{Cai:2001dz}:
\begin{equation}
\label{metric-gb1}
ds^2=-n\frac{r^2}{L^2}f_{GB}(r)dt^2+\frac{r^2}{L^2}d\vec{x}^2+\frac{L^2}{r^2f_{GB}(r)}dr^2 ,
\end{equation}
where
\begin{equation}
\begin{split}
f_{GB}(r)&=\frac{1}{2\lambda_{GB}}\left(1-\sqrt{1-4\lambda_{GB}(1-r_+^4/r^4)}\right)\,,\\
n&=\frac{1}{2}\left(1+\sqrt{1-4\lambda_{GB}}\right)\,.
\end{split}
\end{equation}
The boundary of the metric (\ref{metric-gb1}) is at $r\rightarrow \infty$. We choose the positive parameter $n$ to specify the speed of light of the boundary gauge theory be unity. As a result, one can easily find 
\begin{equation}
    f_{GB}(r\xrightarrow{}\infty)=\frac{1}{n}.
\end{equation}
The heat bath temperature of the black hole is given by
\begin{equation}
T_{GB}=\frac{\sqrt{n}r_+}{\pi L^2},
\end{equation}
where $r_+$ depends on $\lambda_{GB}$ for a fixed Hawking temperature.

Using a general form of metric, one has
\begin{equation}
\begin{split}
    g_{tt}(r)=-n\frac{r^2}{L^2}f_{GB}(r),\qquad
    g_{ii}(r)=\frac{r^2}{L^2}, \qquad
    g_{rr}(r)=\frac{L^2}{r^2f_{GB}(r)}.
\label{metric-gb2}
\end{split}
\end{equation}
In our analysis, we set $AdS_5$-radius to be unity for convenience. Using the same procedures as before, we can easily find the critical value $r^{c}_{GB}$ where the numerator and denominator change sign at the same value,
\begin{equation}
r^{c}_{GB}=\frac{\sqrt{n}(r_+)}{\left(n (n-v^2)+\lambda_{GB}
v^4\right)^{\frac{1}{4}}}.
\end{equation}
The world-sheet temperature of GB gravity of a quark feel is denoted as $T^{ws}_{GB}$,
\begin{equation}
\begin{aligned}
T^{ws}_{GB} = \frac{1}{2\pi} \left(\left| \frac{-4n\lambda_{GB}(r_+)^8 +2n(r^c_{GB})^4 r_+^4K}{\lambda_{GB}(r^c_{GB})^2 \left(\left(-1+4\lambda_{GB} \right)(r^c_{GB})^4-4\lambda_{GB} r_+^4\right)} \right|\right)^{\frac{1}{2}},&\\
K=\left(-1+4\lambda_{GB}+\sqrt{1+\lambda_{GB}\left(-4+\frac{4r_+^4}{(r^c_{GB})^4}\right)} \right).
\end{aligned}
\end{equation}

Asserting (\ref{metric-gb2}) to (\ref{tws}), (\ref{kappal}) and (\ref{kappat}), one gets longitudinal and perpendicular LGV-coefficients as
\begin{equation}
\label{gbkappat}
\begin{split}
\kappa^{\perp}_{GB} &=\frac{\sqrt{\lambda}}{\pi}(r^c_{GB})^2T^{ws}_{GB}
\end{split}
\end{equation}
and
\begin{equation}
\begin{split}
\label{gbkappal}
\kappa^{\parallel}_{GB} &= -\frac{-2\sqrt{\lambda}\lambda_{GB}(r^c_{GB})^4r_+^4+(r^c_{GB})^8K}{2 \pi \lambda_{GB} (r_{GB}^c)^2r_+^4} T^{ws}_{GB}.
\end{split}
\end{equation}
One can check that (\ref{gbkappat}) and (\ref{gbkappal}) reduce to the  results in the conformal limit (\ref{lgvsym}) by taking the limit of $\lambda_{GB} \rightarrow 0$.

\begin{figure}[H]
    \includegraphics[width=8.6cm]{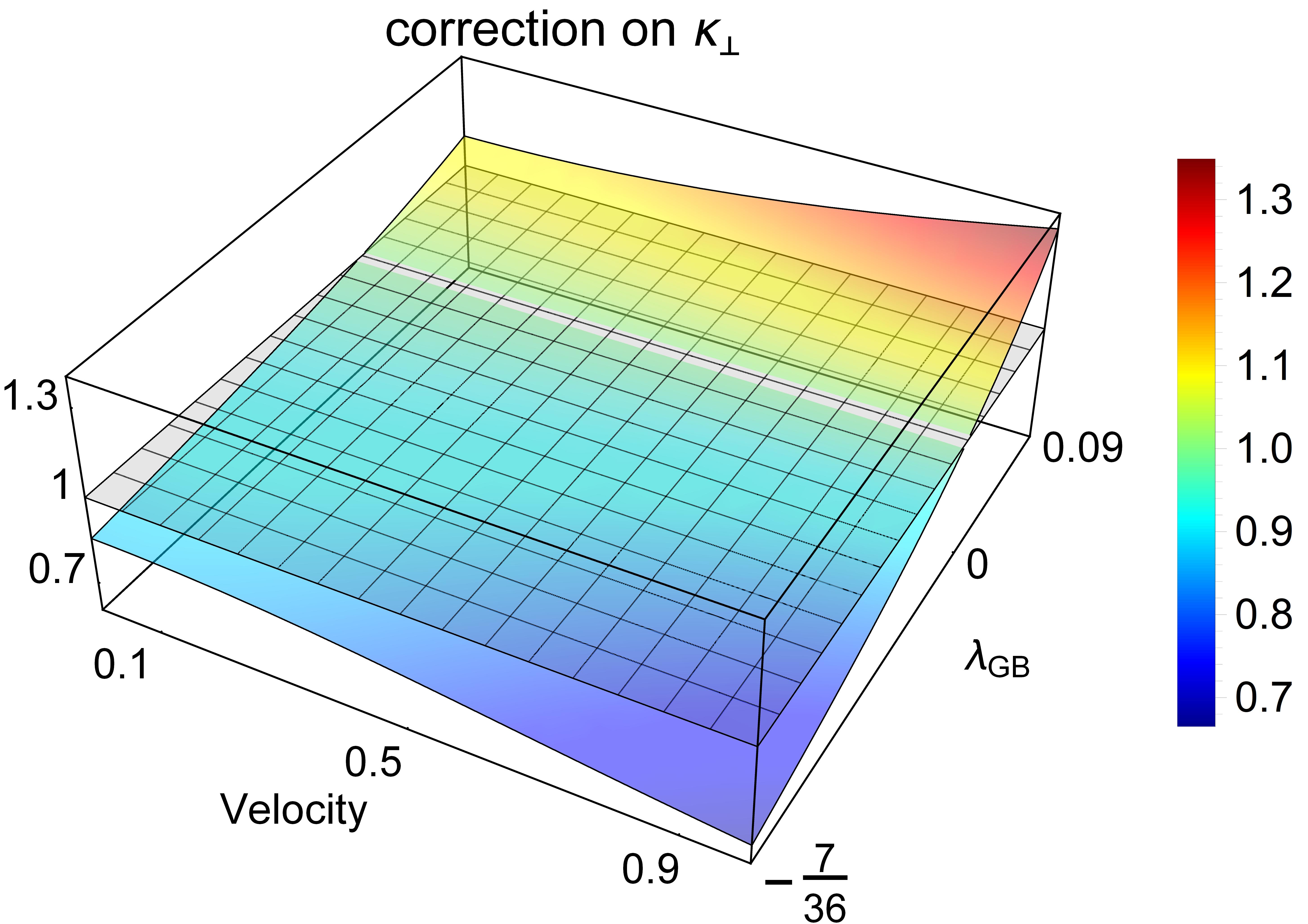}
    \includegraphics[width=8.6cm]{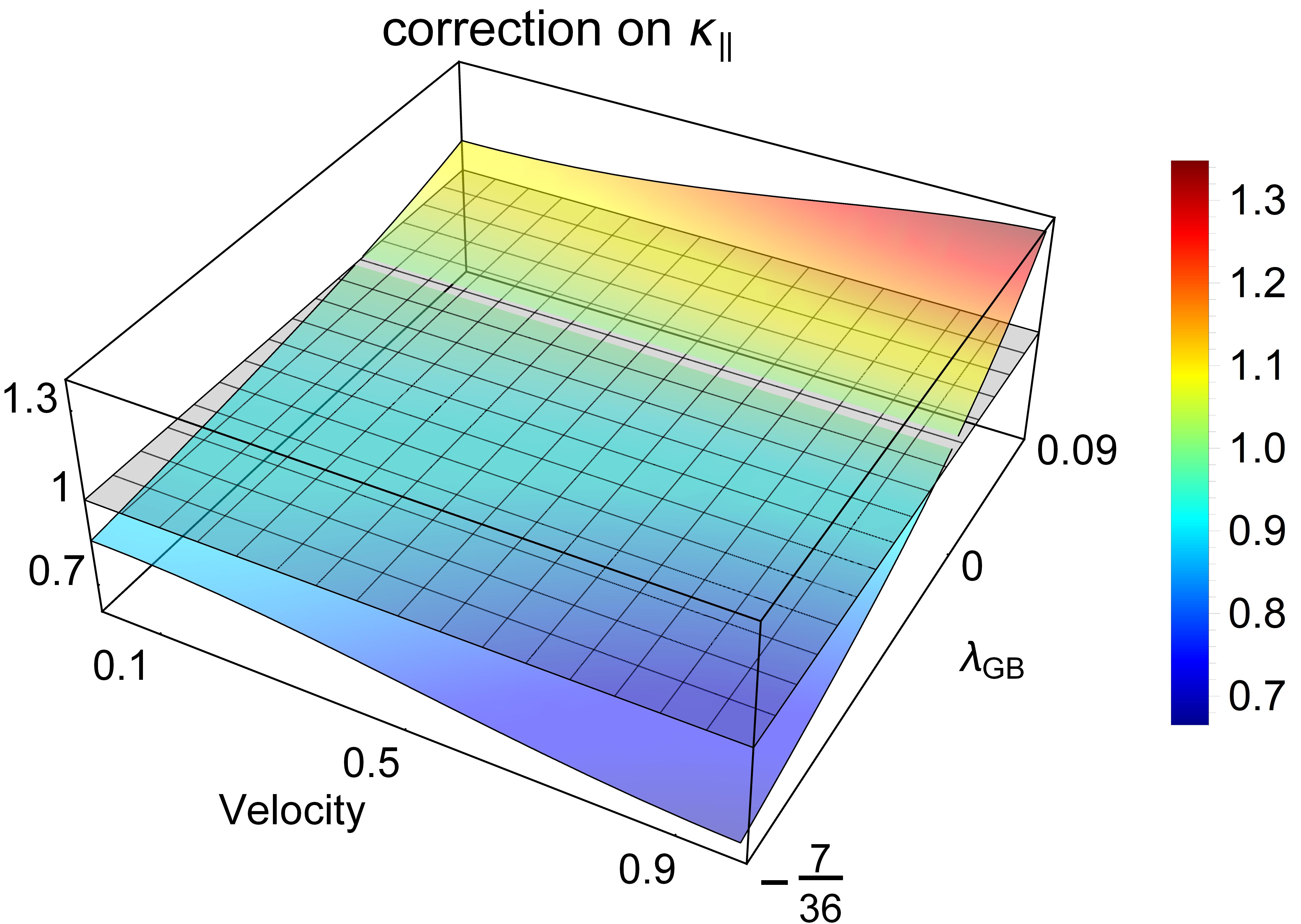}
    \caption{The corrections to the transverse LGV-coefficients $\kappa_{\perp}$ (left panel) and the longitudinal LGV-coefficients $\kappa_{\parallel}$ (right panel) are shown as functions of both velocity and $\lambda_{GB}$, respectively normalized by the conformal limit.}
    \label{figure:EB}
\end{figure}

By employing GB gravity, we now discuss the $\mathcal{R}^2$ corrections on the LGV-coefficients, normalized by the limits given in (\ref{lgvsym}) for $\mathcal{N}=4$ SYM, with two scenarios at the same heat bath temperature ($T_{GB}=T_{SYM}$). 
Plots (a) and (b) in Figure \ref{figure:EB} depict $\kappa_{\perp}$ and $\kappa_{\parallel}$ respectively as functions of moving velocity and $\lambda_{GB}$.
It is evident that both the $\kappa_{\perp}$ and $\kappa_{\parallel}$ are independent of the values of moving velocity and $\lambda_{GB}$. 
It's also found that the correction behaviors to $\kappa_{\perp}$ and $\kappa_{\parallel}$ are notably similar, and the universal relation $\kappa_{\parallel} \geq\kappa_{\perp}$ identified by the authors of \cite{Gursoy:2010aa} also holds in the context of GB gravity.

It is found that the results of LGV-coefficients in the GB gravity, when $\lambda_{GB}=0$, will reduce to the case corresponding to $\mathcal{N}=4$ SYM.
For $\lambda_{GB}>0$, Fig.\ref{figure:EB} demonstrates that the corrections to the $\kappa_{\perp}$ and $\kappa_{\parallel}$ become stronger monotonically with increasing velocity of the moving heavy quark or with the increasing $\lambda_{GB}$. 
Conversely, for $\lambda_{GB}<0$, the $\kappa_{\perp}$ and $\kappa_{\parallel}$ for a moving heavy quark under GB gravity will be less than those in the $\mathcal{N}=4$ SYM case.
Furthermore, the corrections to the $\kappa_{\perp}$ and $\kappa_{\parallel}$ increase monotonically with the increasing the absolute value of $\lambda_{GB}$, and also increase with the growing velocity of the moving heavy quark.
We conclude that the finite coupling corrections affect the both $\kappa_{\perp}$ and $\kappa_{\parallel}$ on a moving quark in the strongly-coupled plasma and depend on the details of curvature squared corrections. The LGV-coefficient can be larger than or smaller than that in the infinite-coupling case. 

\section{Summary}\label{sec4}
Using classical gravity to understand a quantum system is one of the most profound discoveries of contemporary theoretical physics. 
The majority of computations, achieved through classical two-derivative gravity calculations, hold strict validity within the context of a large 't Hooft coupling $\lambda$ and the limit of color number $N_c$.
The modification of quenched jets provides one of the most effective tools for constraining properties of the QGP produced in heavy ion collisions. 
In this study, we take an investigation on finite coupling corrections to heavy quark diffusion. 

We have examined the influences from curvature-squared $\mathcal{R}^2$ corrections on the AdS black brane metric to LGV-coefficients with both a more general $\mathcal{R}^2$ gravity and the GB gravity. 
Our investigation reveals that finite coupling corrections can indeed impact the $\kappa_{\perp}$ and $\kappa_{\parallel}$ values of a moving heavy quark.
Both the $\kappa_{L}$ and $\kappa_{T}$ can be larger than or smaller than that in the infinite coupling case, and the specific correction behaviors depend on the details of higher derivative gravity.
We confirm the persistence of the universal relation $\kappa_{L}\geq\kappa_{T}$ across all cases we examined.
Both corrected $\kappa_{L}$ and corrected $\kappa_{T}$ values can either increase or decrease in comparison to the infinite coupling scenario in GB background.
Our findings regarding curvature squared corrections to $\kappa_{L}$ and $\kappa_{T}$ closely resemble the features obtained for the drag force on a moving heavy quark discussed in \cite{Fadafan:2008gb}.  
Finally, we should emphasize that we do not predict the effects of finite 't Hooft correction to $\mathcal{N} = 4$ SYM, since the leading correction in gauge theory enters at order $R^4$.

\section*{ACKNOWLEDGMENTS}
This research is supported by the Guangdong Major Project of Basic and Applied Basic Research No. 2020B0301030008, and Natural Science Foundation of China with Project No. 11935007. Q. Zhou also expresses gratitude to the China Scholarship Council under Contract No.202306770010. 

\bibliography{ref}
\end{document}